# On the sulfur doping of γ-graphdiyne: A Molecular Dynamics and DFT study

Eliezer Fernando Oliveira[1,2,3], Augusto Batagin-Neto[4], and Douglas Soares Galvao[1,2]

[1]Gleb Wataghin Institute of Physics, University of Campinas (UNICAMP), Campinas, SP, Brazil.
[2]Center for Computational Engineering & Sciences (CCES), University of Campinas (UNICAMP), Campinas, SP, Brazil.
[3]Department of Materials Science and Nanoengineering, Rice University, Houston, TX, United States.
[4]São Paulo State University (UNESP). Campus of Itapeva, Itapeva 18409-010, SP, Brazil.

*Recently, an experimental study developed an efficient way to obtain sulfur-doped γ-graphdiyne. This study has shown that this new material could have promising applications in lithium-ion batteries, but the complete understanding of how the sulfur atoms are incorporated into the graphdiyne network is still missing. In this work, we have investigated the sulfur doping process through molecular dynamics and density functional theory simulations. Our results suggest that the doped induced distortions of the γ-graphdiyne pores prevent the incorporation of more than two sulfur atoms. The most common configuration is the incorporation of just one sulfur atom per the graphdiyne pore.*

## INTRODUCTION:

2D planar carbon materials have been used in a variety of applications, such as rechargeable batteries, capacitors, catalysts, solar cells, and others [1]. Besides the unique intrinsic properties of these materials, they also present a high versatility of synthesis and chemical modifications that allows tuning their electronic and/or mechanical properties. Among the strategies used to achieve this, chemical doping with heteroatoms has been one of the most exploited [2, 3, 4].

In particular, in a recent study, Yang *et al.* [4] reported an effective way of doping γ-graphdiyne (γ-GDY, see Figure 1(a)) with sulfur atoms. In their study, benzyl disulfide (BDS) molecules were mixed with γ-GDY and the temperature of this system was raised to 600° C (873.2 K) to induce the BDS homolysis. This process allows the sulfur atoms to interact with the γ-GDY, thus resulting in a new sulfur-doped γ-graphdiyne (S-γ-GDY). This new material presented improved electrochemical performances in relation to undoped γ-GDY for applications in lithium-ion batteries (LIB) [4]. However, despite the promising features of this system, it is still not fully understood how the sulfur atoms react with the γ-GDY layer. Due to the pore size and the number of triple bonds of the acetylenic linkages in γ-GDY, Yang *et al.* [4] have suggested that it may be possible to have up to three sulfur atoms in each γ-GDY structural unit, as presented in Figure 1(b), but this is only a hypothesis.

To have a better insight of the sulfurization process of γ-GDY, and on the number of incorporated sulfur atoms, we have combined fully atomistic reactive molecular dynamics (MD) and density functional theory (DFT) simulations to address these questions. We have tested the sulfur-doping process at two different temperatures, at room temperature (300 K) and the experimental one (873.2 K). Our MD results show that, as expected, there is a significant dependence on the number of sulfur atoms that react with γ-GDY on the temperature of the system, and the regions at which the sulfur atoms are incorporated have the same trends. In general, our combined MD/DFT results suggest that it is possible to find at maximum two sulfur atoms per pore of γ-GDY due to the pore deformations and small changes induced on the average carbon bond lengths with the temperature.

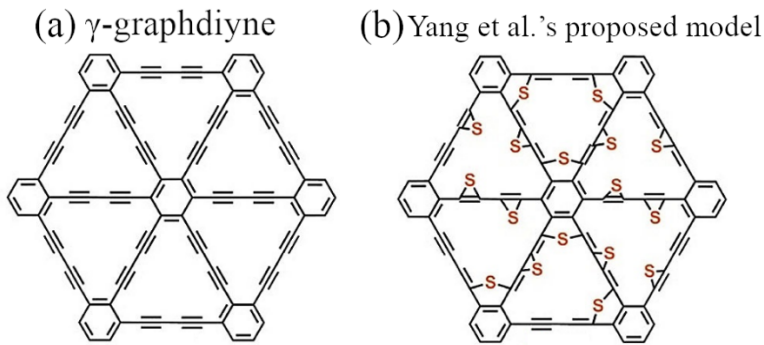

Figure 1. (a) Geometrical structure of γ-graphdiyne (γ-GDY) and (b) Yang *et al.'s* proposed model [4] for the sulfur-doped γ-GDY.

**MATERIAL AND METHODS**

Fully atomistic reactive molecular dynamics simulations (MD) were used to study the γ-GDY sulfur-doping process. First, we created a simulation box with dimensions of 160 x 180 x 140 Å$^3$ and at the position z = 4 Å we placed a γ-GDY sheet containing 7200 carbon atoms, which contains approximately 600 pores (the γ-GDY sheet was previously energy minimized using a conjugated gradient technique). Next, an atmosphere of 1200 sulfur atoms was created (randomly distributed) inside this box. We allowed the x and y direction to be periodic, but we confine the system along the z-direction adding an infinitely hard Lennard-Jones walls at the bottom and top of the box, respectively. Then, we carried out two MD runs of 5.0 ns at 300K and at 873 K, respectively, in an NVT ensemble. The MD timestep used in all simulations was 0.1 fs. These MD simulations were carried out using the reactive force field ReaxFF [5], as implemented in the computational code LAMMPS [6].

To analyze the electronic structure of the γ-GDY at 0 K and 300 K, we also performed DFT simulations, as will be discussed in the next section. The DFT simulations were performed in a projector augmented wave potential (PAW) and a Perdew, Burke, and Ernzerhof (PBE) exchange-correlation functional [7]. The Brillouin zone sampling used a 4 x 4 x 1 Monkhorst-Pack *k*-mesh with a plane-wave basis set and a kinetic energy cutoff of 500 eV. The DFT/PAW/PBE simulations were performed using the computational code Quantum Espresso [8, 9].

## RESULTS AND DISCUSSIONS

In Figure 2(a) we present representative MD snapshots of γ-GDY before (t = 0 ns) and after (t = 5 ns) the sulfur-doping at 300 K; for 873.2 K, the results were similar.

In order to verify whether the sulfur atoms will bond or adsorb into the γ-GDY sheet, we calculated the pair radial distribution (g(r)) between C and S at 300 K and 873.2 K (with r ≤ 4.0 Å), obtained by averaging all of the simulation steps. The results are presented in Fig. 2(b). The two most intense peaks for both temperatures (300 K and 873.2 K) are around 1.93 Å and 2.8 Å. The experimental bond length for C-S bonds is ~ 1.82 Å [10]; DFT and Hartree-Fock methods estimate this bond length between 1.8-1.9 Å [11]. Then, we can attribute the first peak of g(r) in Figure 2(b) to the C-S bonds. We see that there is no much difference in the position of the first peak for both temperatures. The second peak in g(r) is related to the second-neighbor C-S distances in the same pore.

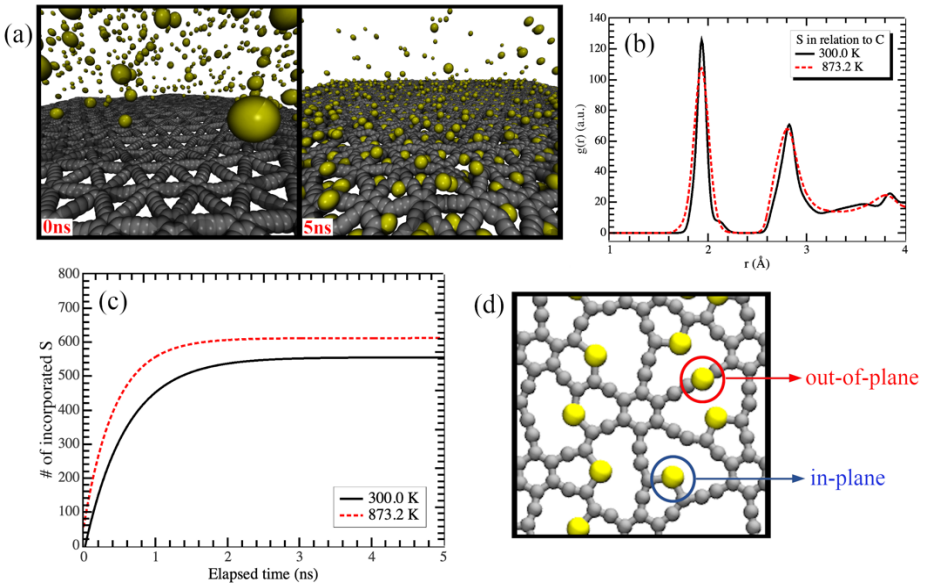

Figure 2. (a) Representative MD snapshots of the γ-graphdiyne before and after the sulfur-doping process, (b) pair radial distribution (g(r)) between C and S at 300 K and 873.2 K, (c) number of incorporated sulfur atoms as a function of the simulation time for 300 K and 873.2 K, and (d) zoomed γ-graphdiyne regions indicating the conformations of the incorporated sulfur atoms.

It is also important to determine the evolution dynamics of the number of incorporated sulfur atoms into tγ-GDY. In Figure 2(c) we present the sulfur evolution as a function of the simulation time for 300 K and 873.2 K. The data presented in Figure 2(c) were fitted using an exponential function $\#S(t) = \#S_0 - Ae^{(-t/B)}$. $\#S(t)$ and $\#S_0$ represent, respectively, the number of sulfur atoms that are incorporated into the γ-GDY in a time t (in ns) and the saturated one (when the incorporation stops); A and B are adjusted parameters. From Figure 2(c) it is clear the temperature effect in the sulfur dynamics, in which the 873.2 K curve remains above that for 300K; this is expected since the temperature affects the speed of sulfur atoms and increases the probability of collision occurrence. According to these exponential fits, we obtained the number of incorporated sulfur atoms (parameter #$S_0$) of ~556 and ~612 for 300 and 873.2 K, respectively. Considering the beginning of saturation in each case, when the number of incorporated sulfur atoms reaches 99% of the corresponding saturated value, the estimated time for saturation are 2.69 and 1.94 ns for 300 K and 873.2K, respectively. These numbers suggest that the temperature increase the reaction velocity, which leads to an earlier sulfur incorporation saturation, as expected.

Regarding the regions in which the sulfur atoms will bond, it is expected that they will occur in the acetylenic linkages of γ-GDY, due to the high electronic density in these regions (triple bonds). Also, according to the experimental work regarding the γ-GDY sulfur-doping, it was suggested that up to three sulfur atoms can remain in each pore. We will now evaluate these points. First of all, the γ-GDY sheets that we are using in our simulations have ~600 pores. According to our simulations, at 873.2 K, 612 sulfur atoms were incorporated into the γ-GDY, which means that in general, we have one sulfur atom per pore, except in some cases (12 remaining sulfur atoms). The sulfur atom needs to make two bonds in order to complete its valence, and our simulation indicates, independently of the temperature, that the majority of them remain bonded to two carbon atoms in adjacent acetylenic linkages, as shown in Figure 2(d). In this case, the sulfur atoms are in-plane and cause a deformation of the pore, which prevents more than one sulfur atom to go into the same pore. There a few of them that remain out of plane bonded to the carbon atoms in the same acetylenic linkage (see Figure 2(d)). Thus, we can conclude that is most probable to have just one sulfur atom per pore of γ-GDY up to the saturation limit.

It is interesting to notice that, there is a preference of the sulfur atom to be bonded to C3 atoms as highlighted in Figure 3(a). As the C2-C3 bonds are part of the triple bonds, we could expect that the sulfur atoms could interact with both of them, and not only with C3 atoms. In order to better understand this point, we calculated the normalized g(r) for the bonds formed between the C1-C1, C1-C2, C2-C3, and C3-C3 highlighted carbon atoms in Figure 3(a) at 0 K and 300 K. The results are presented in Figure 3(b). The g(r) for 0 K was obtained from the geometry optimization of a γ-GDY supercell (with 72 carbon atoms) using a DFT/PAW/PBE approach (see Materials and Methods section); for 300 K, we used the one from the MD simulations. As can be seen in Figure 3(b), there is a modification in the bond lengths of carbon bonds with the temperature, as expected. However, the most interesting case it that the bond length between C3-C3 carbon atoms decrease ~0.1 Å from 0 K to 300 K, presenting a value between the sp and $sp^2$ bond length values. This could modify the charge density, increasing the charge around C3-C3 bonds, making them more favorable to sulfur incorporation. In order to test this hypothesis, we built a γ-GDY structure with the geometry obtained in Figure 3(b) at 300 K and we calculated the resulting charge density from this configuration through a one SCF cycle with a constrained geometry in a DFT/PAW/PBE approach. In Figure 3(c) we present the resulting charge density at 300 K, together with the one obtained from the geometry at 0K. As can be seen, indeed the shortening of the C3-C3 bonds causes an increase of the charge density in the C2-C3 triple bonds. Thus, as the regions where the carbon atoms C3 have a higher electron density then that of carbon atoms C2, there is a preference for the sulfur bonding to the C3 carbon atoms, explaining the results discussed before. Based on these results we can conclude that the most probable configuration of sulfur-doped γ-GDY is the one presented in Figure 3(d). This model was build based on MD results and averaging of the different observed configurations.

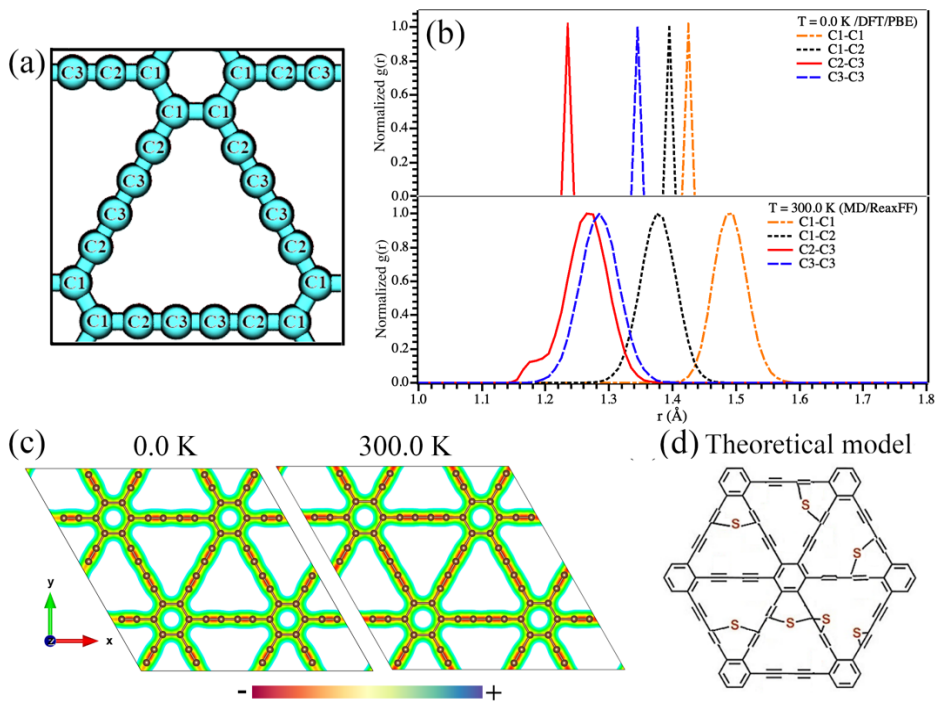

Figure 3. (a) Highlighted atoms that forms different bonds in of γ-graphdiyne pores. (b) Normalized g(r) for the carbon bonds of γ-graphdiyne at 0 K and 300 K. (d) Charge density distribution in of γ-graphdiyne at 0 K and 300 K. (d) Proposed model of the probable resulting structure of the γ-graphdiyne sulfur-doped.

## CONCLUSIONS

In this work, we investigated the dynamics of sulfur-doping γ-graphdiyne combining molecular dynamics (MD) and density functional theory (DFT) methods. Our results suggest that the induced distortions of the γ-graphdiyne pores during the doping process prevent the incorporation of more than two sulfur atoms. The most common configuration is the incorporation of just one sulfur atom per graphdiyne pore, differently from the *Yang et al.*'s proposed model [4], that was 3 sulfur atoms per pore. It should be remarked that we considered pristine γ-graphdiyne structures. If the structures contain defects that results in larger porous, eventually more sulfur atoms could be incorporated due to the decreased steric hindrance among sulfur atoms. More research along these lines is needed to better clarify this issue.

## ACKNOWLEDGMENTS

The authors thank the Brazilian agency FAPESP (Grants 2013/08293-7, 2016/18499-0, and 2019/07157-9) for the financial support and Center for Computational

Engineering and Sciences (CCES) and Center for Scientific Computing of the São Paulo State University (NCC/GridUNESP) for the computational support.

**References:**

1. Y. Sun, Q. Wu, G. Shi, *Energy Environ. Sci.* **4**, 1113 (2011).
2. H. Lee, K. Paeng, I. S. Kim, *Synth. Met.* **244**, 36 (2018).
3. Y. S. Yun, *et al., J. Power Sources* **262**, 79 (2014).
4. Z. Yang, *et al., Chem. Eur. J.* **25**, 5643 (2019).
5. A. C. T. van Duin, S. Dasgupta, F. Lorant, W. A. Goddard, *J. Phys. Chem. A* **105**, 9396 (2001)
6. S. J. Plimpton, *Comput. Phys.* **117**, 1 (1995).
7. J. P. Perdew, K. Burke, M. Ernzerhof, *Phys. Rev. Lett.* **77**, 3865 (1996).
8. P. Giannozzi, *et al.*, *J. Phys.: Condens. Matter.* **21**, 395502 (2009).
9. P. Giannozzi, *et al., J. Phys.: Condens. Matter.* **29**, 465901 (2017).
10. D. R. Lide, *CRC handbook of chemistry and physics*. 88th ed. CRC Press: Boca Raton, 2007.
11. https://cccbdb.nist.gov (acessed in 21/02/2020).